\documentclass[aps,showpacs,twocolumn]{revtex4}
\usepackage{amsmath,amssymb}
\usepackage{graphicx}
\newcommand{\ket}[1]{| #1 \rangle}
\newcommand{\bra}[1]{\langle #1 |}
\newcommand{\brkt}[2]{\langle #1 | #2 \rangle}
\newcommand{\Hil}{\mathcal H}
\DeclareMathOperator{\mymod}{mod}

\newcommand{\Eq}[1]{Eq.~(\ref{#1})}
\begin{document}

\preprint{0708.0145 [quant-ph]}

\title{Programmable quantum state transfer}

\author{Alexander Yu.~Vlasov}
\thanks{E-mails: \raisebox{-3.6pt}[6pt]{\includegraphics{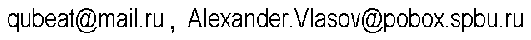}}}
\affiliation{Federal Radiology Center (IRH) \\ 
197101, Mira Street 8, St.--Petersburg, Russia}


\date{\today}

\begin{abstract}
A programmable quantum networks model is used in this paper for 
development of methods of control of a quantum state transport. 
These methods may be applied for a wide variety of patterns of 
controlled state transmission and spreading in quantum systems. 
The programmable perfect state transfer and quantum walk, mobile 
quantum (ro)bots and lattice gas automata may be described by 
unified way with such approach.
\end{abstract}



\maketitle

\section{Introduction}
\label{Sec:Intro}

Different kinds of the non-optical quantum state transport using 
specific phenomena in ``quantum wires'' are investigated very actively
during recent few years. Some references may be found in 
Sec.~\ref{Sec:Transf} and Sec.~\ref{Sec:Walk}. In the present work is 
discussed a compact theoretical approach to the programmable quantum state 
transfer. These methods have applications for the coherent control of
quantum systems, the theory of quantum communications and computations. 
They also uncover promising relations between different models of 
the quantum information science. 

The programmable state transfer is introduced as a particular kind of
conditional quantum dynamics in Sec.~\ref{Sec:Cond}. Application
of these methods to simple motion along lattices controlled by
state of a qubit is discussed in Sec.~\ref{Sec:Qubot}. 
A relation with programmable extension of so-called perfect 
quantum state transfer with lattices and spin chains is considered 
in the  Sec.~\ref{Sec:Transf} and some possibilities of the control of 
the coined quantum walk are described in the Sec.~\ref{Sec:Walk}.

\section{Conditional quantum dynamics}
\label{Sec:Cond}

An essential part of quantum information processing, is the 
``{\em conditional quantum dynamics}, in which one subsystem
undergoes coherent evolution that depends on the 
quantum state of another subsystem'' \cite{BDEJ95}. 

A simple example of the conditional quantum dynamics may be
written as \cite{BDEJ95}
\begin{equation}
 C = \sum_k{\ket{k}\bra{k} \otimes U_k}.
\label{Cond}
\end{equation}

We can denote Hilbert spaces of these two subsystems
as $\Hil_P$ (``program'' or a control system), 
$\Hil_d$ (``data'' or a target system) and
to consider a gate $G$ on $\Hil_P \otimes \Hil_d$.
If we {\em do not accept entanglement between two subsystems},
most general form of conditional or programmable
evolution may be expressed as \cite{NC97}
\begin{equation}
 G : \bigl[\ket{P_U} \otimes \ket{d}\bigr] 
 \to \ket{P'_U} \otimes (U\ket{d}),  
\label{Prog}
\end{equation}
where $\ket{d} \in \Hil_d$ is {\em an arbitrary} state of the target 
system and $\ket{P_U} \in \Hil_P$ is a state of the control system 
(``a program register'') {\em implementing} operator $U$. 

In \cite{NC97} was shown that if two states $\ket{P_\alpha}$ and 
$\ket{P_\beta}$ of the program register implement two different  
operators $U_\alpha$ and $U_\beta$, then \Eq{Prog} implies
\begin{equation}
 \brkt{P_\alpha}{P_\beta} = 0.
\label{OrtProg}
\end{equation}
Due to \Eq{OrtProg} all states of the program register
are orthogonal and the dimension of $\Hil_P$ is equal to the number of
different operators we need to implement. It was used in \cite{NC97}
as an inspiration to the theory of stochastic programmable quantum devices,
but there are also implications to usual unitary evolution, discussed
in the present paper.

Due to \Eq{OrtProg} we may without lost of generality to use states 
of the control register implementing different programs as a new 
computational basis $\ket{k}$ \cite{AV01}. In such a case the operator
$C$ \Eq{Cond} satisfies \Eq{Prog} for the basis states, {\em i.e.},
$
C \bigl[\ket{k} \otimes \ket{d}\bigr] 
 = \ket{k} \otimes (U_k\ket{d}). 
$ 
A possible change of the state of the control system in \Eq{Prog} may be 
described using the composition of $C$ with $A \otimes \openone$, 
{\em i.e.}, an arbitrary unitary operator on the first subsystem.

For an arbitrary state $\ket{\psi} = \sum_k \psi_k \ket{k}$ of the
control register the operator \Eq{Cond} does not satisfy \Eq{Prog}, 
because states of control and target systems become {\em entangled}
\begin{equation}
C \bigl[\ket{\psi} \otimes \ket{d}\bigr] =
 \sum_k{\psi_k\bigl[\ket{k} \otimes (U_k\ket{d})\bigr]}. 
\label{CEnt}
\end{equation}

{\em The programmable quantum state transfer} is defined in present
work as a quantum network ensuring Eqs.~(\ref{Cond},~\ref{Prog})
with {\em spatially distributed system as a target}. 

It should be mentioned, that any universal set of quantum 
gates with distributed quantum systems might include possibility of 
information transfer, {\em e.g.}, quantum interfaces \cite{LLS03} 
intensively use controlled gates with different purposes.
So it is justifiable to restrict discussion here 
to more specialized class of quantum networks described by \Eq{Cond} 
{\em displaying a correspondence principle with standard transport
models like lattice gases and random walks}.

\section{Quantum bots on lattices}
\label{Sec:Qubot}

Perfect cloning of arbitrary unknown quantum states is forbidden \cite{noclon} 
and we may limit consideration to transmission without proliferation. 
A simple model of such a transfer is a linear lattice with $N$ sites, 
two-dimensional control space, and evolution described 
by \Eq{Cond} with $U_0 = U_1^* = U$, where $U$ is the shift operator
\begin{equation}
 B = \ket{0}\bra{0} \otimes U + \ket{1}\bra{1} \otimes U^*,
\quad U_{ij} = \delta_{i,j+1 \mymod N}.  
\label{qubot}
\end{equation}
It may be considered as a rudimentary version of a quantum robot \cite{BQR1},
and a term ``quantum bot'' or ``qubot'' was suggested for such a system
\cite{qubot,QCA}.

In fact, \Eq{qubot} corresponds to quantum mechanical notation
for a simplest example of one-dimensional invertible {\em lattice gas  
cellular automaton} (LGCA) \cite{Inv} with ``control bit'' denoting
direction of motion. A similar model is also known in the theory of 
quantum cellular automata \cite{Mey96} and coined quantum walks
\cite{coin}. The \Eq{qubot} corresponds to 
cyclic boundary condition and similar quantum extensions for boundaries 
with reflections and multidimensional lattices may be also constructed 
using classical LGCA \cite{qubot,QCA,Inv,Mey96}.

It is possible to quantize the ``almost classical'' model of the 
state transfer \Eq{qubot} and adapt it for quantum control or
computations with higher dimensional systems \cite{high,qi02}. For
example, in the quantum case it is possible to consider ``$n$-th roots'' 
of the operator $U$ with properties \mbox{$R^n = U$} and a related
question about {\em continuous generalization} of this discrete 
time model using the limit $n{\to}\infty$ and even about the
Hamiltonian of such evolution.

There is quite straightforward approach to such a question \cite{qubot}
using diagonalization of $U$ by the discrete Fourier transform  
$F_{jk} = e^{2\pi i jk/N}/\sqrt{N}$, {\em i.e.}, a discrete analogue of 
transition between coordinate and momenta spaces
\begin{equation}
 U = F^* V F, 
 \quad V_{kj} =e^{2 \pi i\, k/N}\delta_{kj}.
\label{FUF}
\end{equation}
The matrices $V$ and $U$ form so-called Weyl pair and in 
continuous limit $U$ corresponds to translation operator 
$\exp(i\tau p): \psi(x)\to\psi(x+\tau)$ \cite{Weyl}.

Due to \Eq{FUF} it is possible to suggest an expression for family
of ``roots'' $U(\alpha)$, 
\begin{equation}
 U(\alpha) = F^* V(\alpha) F, \quad 
 V_{kj}(\alpha) =e^{2 \pi i \, \alpha k/N}\delta_{kj}
\label{URoot}
\end{equation}
where $U(\alpha)U(\beta) = U(\alpha+\beta)$, $U(1)=U$ and $R^n=U$
for $R = U(1/n)$. A family of operators $U(\alpha)$ and
$V(\beta)$ resembles {\em the Weyl system} for continuous case \cite{AQFT},
but satisfy necessary Weyl commutation relations only for an integer
$\alpha$, $\beta$ and should be discussed elsewhere. It is also 
possible to suggest a Hamiltonian $H_U$ for the gate $U$  
\begin{equation}
 H_U = F^* K F,\quad 
 K_{kj} = 2\pi k\delta_{kj}/N,\quad 
 e^{i H_U t} = U(t).
\label{UHam}
\end{equation}
Hamiltonians for networks with more general topology are discussed in \cite{KNJ07}.
Coefficients of the matrices \Eq{URoot} and the Hamiltonian \Eq{UHam}
may be simply calculated directly \cite{qubot,sum}
\begin{equation}
 U_{jk}(\alpha) = \mathring{s}_N(k-j+\alpha),
 \quad \mathring{s}_N(x) \equiv 
 \frac{e^{i\varpi x}\sin(\pi x)}{N \sin(\pi x / N)},
\label{Uroot}
\end{equation}
where $\varpi \equiv \pi (N-1)/ N,$
\begin{equation}
 (H_U\!)_{jj} = \varpi,  \quad
 (H_U\!)_{jk} = \frac{2 \pi/N}{1-e^{2\pi i (k-j)/N}},\quad j\ne k.
\label{Uham}
\end{equation}

Using \Eq{UHam}, a Hamiltonian of the conditional evolution \Eq{qubot} may
be represented as
\begin{equation}
 H_\bot = \ket{0}\bra{0} \otimes H_U - \ket{1}\bra{1} \otimes H_U =
 \sigma_z \otimes H_U.
\label{Hs}
\end{equation}

Similar models may switch over different outputs in 
according to a state of control. Let us consider two-dimensional 
control space $\Hil_P$, two coupled lattices with Hilbert space 
$\Hil_{2N} = \Hil_2 \otimes \Hil_N$ as a target,
and the Hamiltonian
\begin{equation}
 H_{\doteq} = \openone \otimes (\openone \otimes H_U) + 
 \tfrac{2}{N-1}\, \ket{1}\bra{1} \otimes (\sigma_x \otimes \openone) 
\label{Hs2}
\end{equation}
on $\Hil_P \otimes (\Hil_2 \otimes \Hil_N)$. It corresponds to 
transmission along one lattice for state $\ket{0}$ of control qubit 
with additional switch between two lattices for $\ket{1}$. 

The \Eq{Hs2} is a degenerated case of ``chessman'' Hamiltonian for 
two-dimensional $m \times n$ lattice \cite{qubot}, {\em i.e.},  
$H_\divideontimes = 
\sum_{k,j}{\ket{k,j}\bra{k,j} \otimes(k \openone \otimes H_U 
 + j H_U \otimes \openone)}$, 
where $(k, j)$ are directions of moves.

The \Eq{Uham} describes a Hamiltonian with long-range interaction and 
the attenuation law approximately proportional to $|j-k|^{-1}$ for 
$N \gg |j-k|$. Such a Hamiltonian may produce some problem with precise 
experimental engineering. It would be good to find an equivalent 
Hamiltonian with nearest-neighbour interaction and it is discussed in 
the next section.

The \Eq{Uroot} shows, that only for integer $\alpha=n$ state
is localized $U(n)\ket{0}=\ket{n}$, contrary to a nonlocal
distribution $U(\alpha)\ket{0}=\sum_k\mathring{s}_N(\alpha-k)\ket{k}$ for 
real $\alpha$.

\section{Perfect state transfer}
\label{Sec:Transf}

In \cite{ChrEk} was suggested a Hamiltonian for the quantum spin chain
for the perfect state transfer. Let us show, that up to change of basis
it produces the same evolution as $H_U$ in \Eq{UHam}. It is known \cite{FLP3}, 
that the Hamiltonian for a Heisenberg chain used in \cite{ChrEk} is 
immediately related with a Hamiltonian for one particle on a 
lattice with tunnelling between neighbor sites,  {\em i.e}, 
with a basic model of the present paper. 

Similar lattice and graph analogues of the chain networks 
are also well known in the quantum information science \cite{tri}. 
Here is suggested for simplicity, that for the
transport of a qubit state is used dual-rail encoding 
\cite{ChrEk,dual}, because such a case has more direct
relation with lattice models used here. 

The operator $K$ in \Eq{UHam} is equal to 
$J_z/h + \varpi \openone$, where $J_z$ corresponds to 
a ``fictitious'' particle with spin $(N-1)/2$,
$h$ is Plank's constant, and $\varpi$ was introduced after \Eq{Uroot}. 
Operators $J_x$, $J_z=(K-\varpi \openone)h$ and 
$H'_U = (H_U-\varpi \openone)h$ have the same eigenvalues and there 
is some operator $O_x$:
$O_x H'_U O_x^* = J_x$. It is enough to use the composition of the
Fourier transform and the transition between $J_x$ and $J_z$ basis 
(see \cite{FLP3,Hamer}).

The operator $\Omega J_x$ with a strength parameter $\Omega$
corresponds to the Hamiltonian of {\em the perfect state transfer} 
introduced in \cite{ChrEk} and resolves the problem with 
a ``nearest-neighbour representation'' of the Hamiltonian 
$H_U$ for the shift operator. 
In this basis instead of the shift matrix $U$ we have 
higher dimensional representation of a rotation
$R_x(t)=\exp(i\, t J_x/\hbar)$ \cite{ChrEk,FLP3,Hamer,Mar}. 
Unlike the operator $\exp(i H_U t)=U(t)$, it displays localization of 
the initial state $\ket{0}$ only for the extreme points of a lattice (chain),
but it is enough for the perfect state transfer.

A conditional Hamiltonian like \Eq{Hs} for such a 
transfer is $\sigma_z \otimes J_x$.
More useful is analogue of \Eq{Hs2}, {\em i.e.},  
$H_{\doteq} = \openone \otimes \openone \otimes J_x + 
  \frac{2 h}{N-1}\ket{1}\bra{1} \otimes \sigma_x \otimes \openone$, 
controlling of switch between two different output lines. 
Such Hamiltonians describe a controlled scalar excitation, but 
for transfer of a qubit state it is enough to double number of 
lattices.

In general, we can consider such a kind of models as a tensor product 
of three Hilbert spaces $\Hil_S \otimes \Hil_P \otimes \Hil_d$, 
{\em viz}, transmitted state, program and distributed target respectively. 
It is also can be considered
as an extension of a control system to $\Hil_S \otimes \Hil_P$, when
only subsystem $\Hil_P$ may affect on transfer by $\Hil_d$. 

Most methods discussed here may be used almost without change both 
for lattices and spin chains, because for the state transfer via spin 
chains with $n$ nodes nowadays \cite{ChrEk,Mar,dual,mult} 
often is used only $n$-dimensional 
subspace of whole $2^n$-dimensional Hilbert space and a lattice 
with $n$ nodes may be used instead.

For a spin chain a simple relation with a lattice model exists
only for spin-half particles and so, using one
lattice with an internal space $\Hil_I = \Hil_S \otimes \Hil_P$ for 
the control and the transferred state, we lose the analogy 
with spin chains. 

It is possible to realize the control with some 
quantum system attached to a single lattice 
or use multiple parallel lattices or spin chains \cite{mult}.
The design with spin chains should utilize some interactions \cite{Mar,BB} 
for conditional dynamics. 
Realistic models of quantum information devices appropriate for 
such purposes may be found in many papers, from earliest 
suggestions \cite{Ll93,Is94} till more recent works \cite{BB,fitz}. 

It should be emphasized, that there is some difference between a model 
of a global control of such chains \cite{fitz} and the programmable
dynamics discussed in the present paper. It is usual
distinction between general and programmable quantum networks 
\cite{Hybr,Nova}, between the external control and the transfer 
driven by an internal state encoding a program of motion.

\section{Coined quantum walk}
\label{Sec:Walk}

A coined quantum walk on a circle \cite{coin} may be considered 
{\em formally} as a special example of conditional quantum dynamics 
$B$ \Eq{qubot} with a control register, altered on each step by the
Hadamard transform $H = (\openone+i\sigma_y)/\sqrt{2}$ 
(or symmetric analogue $(\openone+i\sigma_x)/\sqrt{2}$ \cite{kem}) 
with $T$ steps of evolution described by the operator $(BH)^T$.

The theory of coined quantum walk has interesting outcome to analysis
of the programmable quantum networks, because produces a wide set of
examples with feeding a control register by nonorthogonal states. 
It was mentioned, that for usual theory of (non-stochastic) 
programmable quantum networks \cite{BDEJ95,NC97,AV01,Hybr,Nova} 
different states of a program should be orthogonal to ensure \Eq{Prog} 
and prevent entanglement between the program and a data \Eq{CEnt}.

It is convenient also to compare the coined quantum walk with 
the {\em programmable quantum processors} \cite{AV01,Hybr,Nova}
containing third system, ``a tape'' and a gate $F$ for altering of 
a state of a control register after each step of evolution.
So, instead of one gate $G$ \Eq{Prog} is used an analogue 
of classical processor timing $(FG)^T$ \cite{AV01,Hybr,Nova}. 

A similar design may be used for programmable implementation of 
a coined quantum walk controlled by altered coin(s). 
Different models with set of (random) coins provide possibility 
of ``tuning'' from the quantum to classical-like behavior 
\cite{BCA1,KBH06}. The programmable implementation of such a model
could be compared with the generation of (pseudo)random numbers and
the Monte-Carlo simulations by a classical computer. 

Let us recall a quantum bot $B$ \Eq{qubot} with the control 
register used as a coin space. An application $(BC_\theta)^T$ with coins 
like $C_\theta = \exp(i\theta\sigma_x )$, $\theta \in [0, 2\pi]$ 
provide a smooth transition between an uniform motion and behavior 
of quantum walk \cite{ken}. 
It is possible to introduce a simplest controlled coined quantum walk 
with $n$ different coins $U_k$ and three quantum systems
with a step composed from two operators: $S_{(123)} = C_{(12)} B_{(23)}$.

Here $C_{(12)} \equiv C \otimes \openone$ is $C$ operator \Eq{Cond} for first 
and second systems, $B_{(23)} \equiv \openone \otimes B $ is the operator $B$ 
\Eq{qubot} on second and third systems.
If first system has a state $\ket{k}$ during $T$ steps, $(S_{(123)})^T$ 
is a ``qubot driven'' quantum walk on third system with a coin $U_k$.

A state of a coin may be entangled with a state of a lattice. 
In such a case the second and third systems may be considered
as a joint target, controlled by a state of first system.

Generalization of $C$ \Eq{Cond} for continuous parameters may 
be produced by the simple change of the sum to an integral \cite{Hybr,Nova} 
and let us use smooth tuning of coins like $C_\theta$ above.

\section{Conclusions}

In this work was considered unified approach to different models of the 
programmable quantum state transfer. It was used some methods
of construction of programmable quantum networks with 
a higher-dimensional target system adapted for specific properties
of distributed dynamical models.

It was shown, that a simple ``qubot'' model may be extended to
a programmable system associated with a short-range Hamiltonian, 
coinciding with $\Omega J_x$ operator for some fictitious particle with 
high spin and widely used nowadays in the theory of the perfect state 
transfer.

In the paper was considered only the {\em coined} quantum walk because 
of particular structure. Formally, the coin space resembles 
a specific version of a control register and so a programmable model of 
such a system should use a ``cascade'' with two control registers for
a single target system. It may be formally treated also using
a joint system with a coin and a lattice as a new target for control.

It is shown also, that the application of the theory of programmable 
quantum networks illustrates some useful relations between three 
models mentioned above: the quantum bots and lattice gas automata
(see Sec.~\ref{Sec:Qubot}), the perfect quantum state transfer 
(see Sec.~\ref{Sec:Transf}), and the coined quantum walk 
(see Sec.~\ref{Sec:Walk}).



\begin{thebibliography}{10}
\newcommand{\tit}[1]{``#1,''} 
\newcommand{\arx}[1]{{\em Preprint\/} {\tt #1}} 
\newcommand{\jpa}{J.\ Phys.\ A: Math.\ Gen.\ } 
\bibitem{BDEJ95} A. Barenco, D. Deutsch, A. K. Ekert, and R. Jozsa,
 \tit{Conditional quantum dynamics and logic gates}
 {\em \prl \bf 74}, 4083--4086 (1995). 
\bibitem{NC97} M. A. Nielsen and I. L Chuang, \tit{Programmable quantum
 gate arrays} {\em \prl}{\bf 79}, 321--324 (1997);
\arx{quant-ph/9703032}. 
\bibitem{AV01} A. Yu. Vlasov, \tit{Classical programmability is enough for 
quantum circuits universality in approximate sense} \arx{quant-ph/0103119}.
\newblock A. Yu.\ Vlasov, \tit{Universal quantum processors with 
  arbitrary radix $n{\ge}2$} {\em Int.\ Conf.\ Quantum 
  Inf., ICQI 2001}, ?--?; \arx{quant-ph/0103127}. 
\bibitem{LLS03}S. Lloyd, A. J. Landahl, and J.-J. E. Slotine,
\tit{Universal quantum interfaces}
{\em \pra} {\bf 69}, 012305 (2004);
\arx{quant-ph/0303048}.
\bibitem{noclon} W. K. Wootters and W. H. Zurek,
\tit{A single quantum cannot be cloned}
{\em Nature} {\bf 299}, 802--803 (1982).
\bibitem{BQR1} P. Benioff, \tit{Quantum robots and environments}
{\em \pra \bf 58}, 893--904 (1998); \arx{quant-ph/9802067}. 
\bibitem{qubot} A. Yu.\ Vlasov, ``Quantum `bots' on lattices,''
(Workshop ``Think-tank on Computer Science Aspects
[of Quantum Computation],'' Institute for Scientific
Interchange, Torino, Italy, 19--30 June 2000), unpublished.
\bibitem{QCA} A. Yu.\ Vlasov, \tit{On quantum cellular automata}
\arx{quant-ph/0406119} (2004).
\bibitem{Inv} T. Toffoli and N. Margolus, \tit{Invertible cellular
automata: A review} {\em Physica D \bf 45}, 229--253 (1990).
\bibitem{Mey96}D. A. Meyer, \tit{From quantum cellular automata to 
quantum lattice gases} {\em J. Stat. Phys. \bf 85}, 551--574 (1996);
\arx{quant-ph/9604003}. 
\bibitem{coin} D. Aharonov, A. Ambainis, J. Kempe, and U. Vazirani,
\tit{Quantum walks on graphs} in {\em Proc. 33rd STOC}, 
(ACM, New York, 2001) 50--59; \arx{quant-ph/0012090}.
\bibitem{high} D. Gottesman, \tit{Fault-tolerant quantum computation with
higher-dimensional systems}
{\em Lect.\ Not.\ Comp.\ Sci.\ \bf 1509}, 302--313 (1999);
\arx{quant-ph/9802007}.
\bibitem{qi02} A. Yu.\ Vlasov, \tit{Algebra of quantum computations with 
higher dimensional systems} {\em Proc.\ SPIE \bf 5128}, 
29--36 (2003);
\arx{quant-ph/0210049}.
\bibitem{Weyl} H. Weyl, {\em The Theory of Groups and Quantum
Mechanics}, (Dover Publications, New York, 1931).
\bibitem{AQFT} N. N. Bogoliubov, A. A. Logunov, A. I. Oksak, and 
I. T. Todorov, {\em General Principles of Quantum Field Theory}, 
(Nauka, Moscow, 1987; Kluwer, Dordrecht, 1990).
\newblock J.~C. Baez, I.~E. Segal, and Z. Zhou, {\em Introduction to
Algebraic and Constructive Quantum Field Theory}, (Princeton University
Press, Princeton, 1992). 
\bibitem{KNJ07} V. Ko\v{s}t\!'\!\'ak, G. M. Nikolopoulos, and I. Jex, 
\tit{Perfect state transfer in networks of arbitrary topology and
coupling configuration} 
{\em \pra \bf 75}, 042319 (2007);
\arx{quant-ph/0702016}.
\bibitem{sum} L. B. W. Jolley, {\em Summation of Series}, 
(Dover Publications, New York, 1961).
\bibitem{ChrEk} M. Christandl, N. Datta, A. Ekert, and A. J. Landahl,
\tit{Perfect state transfer in quantum spin networks}
{\em \prl \bf 92}, 187902 (2004); \arx{quant-ph/0309131}.
\newblock M. Christandl, N. Datta, T. C. Dorlas, A. Ekert, A. Kay, 
and A. J. Landahl, \tit{Perfect transfer of arbitrary states in quantum 
spin networks} {\em \pra \bf 71}, 032312 (2005); \arx{quant-ph/0411020}.
\bibitem{FLP3} R. P. Feynman, R. B. Leighton, and M. Sands,
{\em The Feynman Lectures on Physics. III. Quantum Mechanics},
(Addison-Wesley, Reading, MA, 1965). 
\bibitem{Hamer}M. Hamermesh, {\em Group Theory and its Applications to
Physical Problems}, (Addison-Wesley, Reading, MA, 1964).
\bibitem{Mar} A. Kay and M. Ericsson, \tit{Geometric effects and 
 computation in spin networks} {\em New J.\ Phys.\ \bf 7}, 
143 (2005); \arx{quant-ph/0504063}. 
\bibitem{tri} E. Farhi and S. Gutmann, \tit{Quantum computation and 
decision trees} {\em \pra \bf 58}, 915--928 (1998); \arx{quant-ph/9706062}.
\bibitem{dual} D. Burgarth and S. Bose, \tit{Conclusive and arbitrarily 
perfect quantum state transfer using parallel spin chain channels}
{\em \pra \bf 71}, 052315 (2005); \arx{quant-ph/0406112}.
\bibitem{mult} D. Burgarth, V. Giovannetti, and S. Bose, 
\tit{Efficient and perfect state transfer in quantum chains}
{\em \jpa \bf 38}, 6793--6802 (2005); \arx{quant-ph/0410175}.
\bibitem{BB}S. C. Benjamin, \tit{Schemes for parallel quantum computation
without local control of qubits} 
{\em \pra \bf 61}, 020301(R) (2000); \arx{quant-ph/9909007}.
\newblock S. C. Benjamin and S. Bose, \tit{Quantum computing 
with an always-on Heisenberg interaction} {\em \prl \bf 90}, 247901 (2003);
\arx{quant-ph/0401071}. 
\newblock S. C. Benjamin, \tit{Multi-qubit gates in arrays 
coupled by `always on' interactions} {\em New J.\ Phys.\ \bf 6}, 61 (2004);
\arx{quant-ph/0403077}.
\bibitem{Ll93} S. Lloyd, \tit{A potentially realizable quantum computer} 
{\em Science} {\bf 261}, 1569--1571 (1993).
\bibitem{Is94} G. Berman, G. Doolen, D. Holm, and V. Tsifrinovich, 
\tit{Quantum computer on a class of one-dimensional Ising systems} 
{\em Phys.\ Lett.\ A} {\bf 193}, 444--450 (1994).
\bibitem{fitz} J. Fitzsimons and J. Twamley, \tit{Globally controlled 
quantum wires for perfect qubit transport, mirroring, and computing}
{\em \prl \bf 97}, 090502 (2006); \arx{quant-ph/0601120}.  
\newblock J. Fitzsimons, L. Xiao, S. C. Benjamin, and J. A. Jones, 
\tit{Quantum information processing with delocalized qubits under global
control} {\em \prl \bf 99}, 030501 (2007); \arx{quant-ph/0606188}.
\bibitem{Hybr} A. Yu.\ Vlasov, \tit{Universal hybrid quantum processors}
{\em Part.\ Nucl.\ Lett.\ \bf 1[116]}, 60--65 (2003); \arx{quant-ph/0205074}.
\bibitem{Nova} A. Yu.\ Vlasov, ``Programmable quantum networks with pure 
states,'' in J.~E. Stones (ed), {\em  Computer Science and Quantum Computing},
(Nova Science Publishers, New York, 2007), pp. 33--61;
\arx{quant-ph/0503230}.
\bibitem{kem} J. Kempe, \tit{Quantum random walks --- an introductory overview}
{\em Contemp.\ Phys.\ \bf 44}, 307--327 (2003); \arx{quant-ph/0303081}.
\bibitem{BCA1}T. A. Brun, H. A. Carteret, and A. Ambainis,
\tit{Quantum to classical transition for random walks}
{\em \prl \bf 91}, 130602 (2003); \arx{quant-ph/0208195}.
\bibitem{KBH06} J. Ko\v{s}\'{\i}k, V. Bu\v{z}ek, and M. Hillery, \tit{Quantum 
walks with random phase shifts} {\em \pra \bf 74}, 022310 (2006); 
\arx{quant-ph/0607092}. 
\bibitem{ken} V. Kendon, \tit{Decoherence in quantum walks --- a review}
{\em Math.\ Struct.\ in Comp.\ Sci.\ \bf ?}, ? -- ? (200?);
\arx{quant-ph/0606016}.
\end{thebibliography}
\end{document}